\documentclass[aps,prl,twocolumn,groupedaddress,showpacs]{revtex4}
\usepackage{graphicx,amsmath,amssymb,amsfonts,latexsym,color,dcolumn,bm}
\usepackage{verbatim}

\begin{document}

\title{Switching of $\pm$$360^\circ$ domain wall states in a nanoring by an azimuthal Oersted field}

\author{Nihar R. Pradhan$^{1,2}$, Abigail~Licht$^1$, Yihan~Li$^1$, Yineng Sun$^1$,  Mark T.~Tuominen$^2$ and Katherine E.~Aidala$^1$}

\affiliation{ $^1$ Department of Physics, Mount Holyoke College, South Hadley, MA 01075, USA\\
  $^2$ Department of Physics, University of Massachusetts, Amherst, MA 01003, USA}

\begin{abstract}
We demonstrate magnetic switching between two $360^\circ$ domain wall vortex states in cobalt nanorings, which are candidate magnetic states for robust and low power MRAM devices.  These $360^\circ$ domain wall (DW) or ``twisted onion" states can have clockwise or counterclockwise circulation,  the two states for data storage.  Reliable switching between the states is necessary for any realistic device.  We accomplish this switching by applying a circular Oersted field created by passing current through a metal atomic force microscope tip placed at the center of the ring.  After initializing in an onion state, we rotate the DWs to one side of the ring by passing a current through the center, and can switch between the two twisted states by reversing the current, causing the DWs to split and meet again on the opposite side of the ring.  A larger current will annihilate the DWs and create a perfect vortex state in the rings.

\end{abstract}



\maketitle

\section{Introduction}
\begin{figure}[tb]
\includegraphics[width=9.0cm]{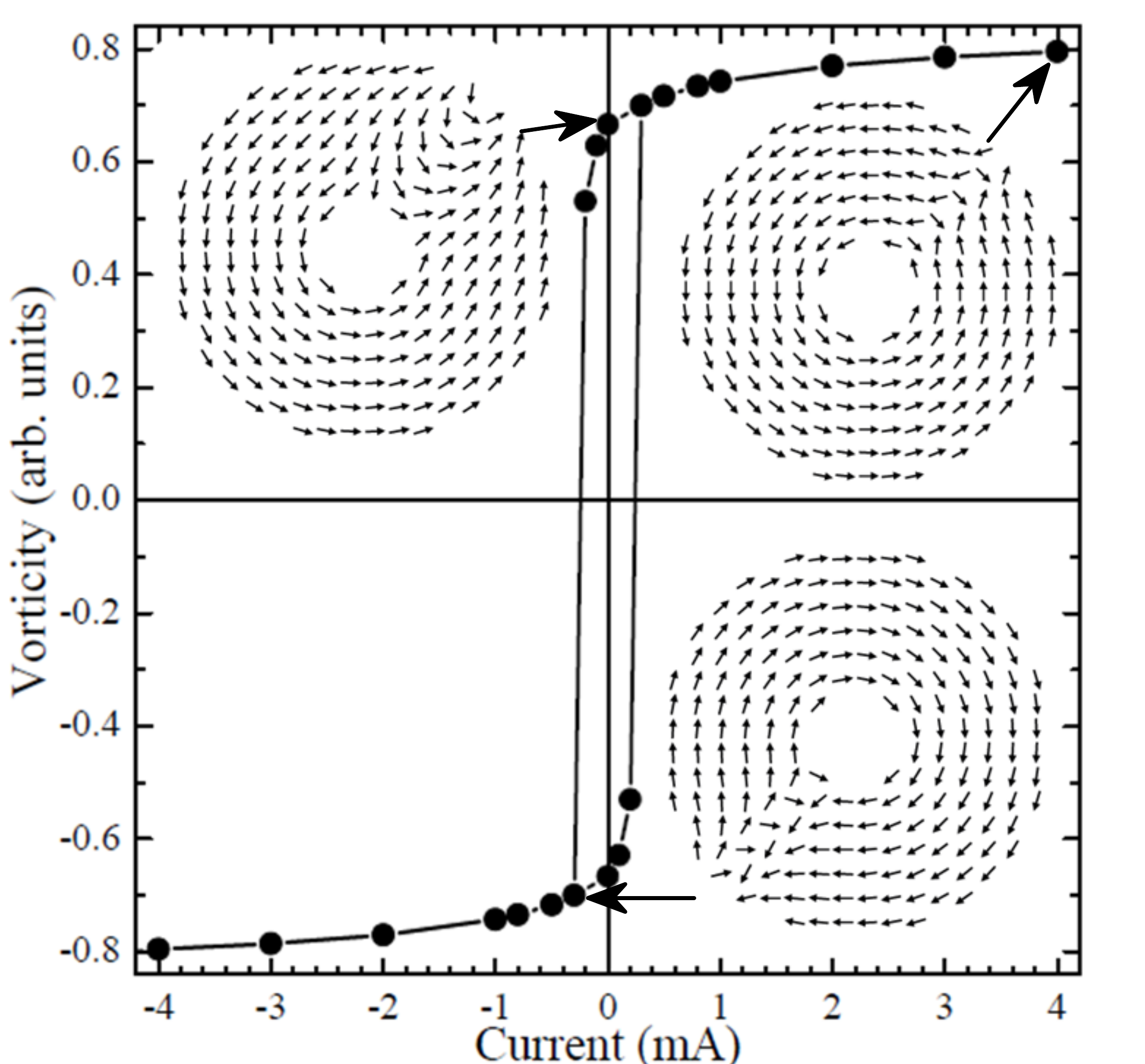}
\caption{\label{unpeel} 
Simulated hysteresis of a 800~nm/200~nm (outer diameter/inner diameter) 5~nm thick permalloy ring as a current is applied through the center of the ring.  The ring is initialized in a CCW $360^\circ$ DW state (upper right inset).  As the current is lowered and reversed, the DW widens (upper left inset) and unpeels, reversing the vorticity of the ring (lower right inset).}
\end{figure}

Magnetoresistive random access memory (MRAM) could serve as a non-volatile random access memory, replacing hard drives, DRAM, and SRAM  if a suitably robust, low power, and dense element can be developed.  A variety of proposals exist with some commercial realization~\cite{Zhu08}.  One promising candidate is to use magnetic nanorings, with the storage bit consisting of the vortex state in which the moments align circumferentially in the clockwise (CW) or counter-clockwise (CCW) direction~\cite{Zhu00}.  The ring geometry is relatively insensitive to small geometric variations and the vortex state has no stray field, leading to robust switching characteristics and dense packing.  The initial proposal~\cite{Zhu00} suffered from requiring relatively high currents to switch fully from one vortex circulation to the opposite.  The current can be reduced by using a slightly different state, called a twisted onion or a $360^\circ$DW state, in which the ring is primarily in the clockwise or counterclockwise vortex, but a single $360^\circ$DW remains \cite{Zhu03, Zhu08, Muratov09} (see Fig.~\ref{unpeel}). Spin torque transfer can further reduce this switching current \cite{Zhu08,Zhu03}. Further reported results suggested the spin torque efficiency decreases as the temperature increases, though the applied field required to switch states is lowered~\cite{Laufenberg06}.

Ferromagnetic nanorings have attracted attention in the research community, in part due to the unique closed-flux vortex state \cite{Klaui03,Klaui07,Zhu00,Zhu08,Zhu06}. Considerable work on simple single-layer structures allowed direct imaging of the magnetization, but  manipulating the magnetic states is limited to uniform external fields, making the control of the vortex circulation challenging.  Introducing asymmetry to the ring enables control over the circulation with a uniform applied field~\cite{Saitoh04}.

In this report, we demonstrate proof of principle control over switching between a twisted onion or $360^\circ$DW vortex state and the opposite circulation state by passing a current through the center of a ring using an atomic force microscope (AFM) tip and imaging the resulting state with magnetic force microscopy (MFM).  The vortex circulation is determined by the Oersted field created by the current through the tip.  Previously, we have demonstrated motion of $180^\circ$DWs and direct vortex to vortex switching in permalloy nanorings \cite{Pradhan11}.  Here we are able to ``unpeel" a $360^\circ$DW and recombine the two $180^\circ$ DWs on the opposite side of the ring, demonstrating the switching between  states that will lower the switching current \cite{Zhu03, Muratov09} for MRAM devices.

Direct switching of one vortex circulation to the opposite in thin and narrow nanorings requires DW nucleation and motion.  Nucleating a DW requires larger external field than coherent rotation or movement of a DW.  An alternative method to switching the vortex circulation in thin and narrow nanorings is shown in Fig.~\ref{unpeel}. Figure~\ref{unpeel} plots the simulated hysteresis, showing vorticity vs. current, simulated using OOMMF~\cite{OOMMF}.  We apply a magnetic field that results from a current through the center of the ring, modeled as an infinite wire with a diameter of 100~nm, and calculate the resulting vorticity of the magnetization.  Vorticity is the degree to which the ring is in a perfect vortex state, defined by
\begin{equation}
V = \frac{1}{A} \int_A (\hat{r} \times \vec{M}) \cdot \vec{dA}
\end{equation}
where $A$ is the area of the ring and a vorticity of $+1$ corresponds to a perfect CCW vortex, and $-1$ to a perfect CW vortex. The ring is in the CCW state with a single $360^\circ$ DW at +4~mA, with the corresponding magnetization shown in the upper right inset of Fig.~\ref{unpeel}. As the current is reduced to zero, the DW widens, with the magnetization shown at remanence in the upper left inset.  After flipping the direction of the current, the DW ``unpeels" at -0.3~mA into the two $180^\circ$ DWs that made up the $360^\circ$ DW, recombining to form a $360^\circ$ DW again on the opposite side of the ring, switching into opposite circulation vortex.  The corresponding magnetization of this state is shown in the lower right inset of Fig.~\ref{unpeel}.  The predicted switching current of 0.3~mA is for a permalloy nanoring with outer diameter 800~nm, inner diameter 200~nm, and a thickness of 5~nm. This method for switching, proposed by Muratov et al.~\cite{Muratov09}, does not require nucleation or annihilation of a domain wall and occurs at a smaller current than direct vortex to vortex switching~\cite{Gabriel09}. A similar method using current through the ring material itself (resulting in a smaller field for the same current, but also enabling spin torque transfer contributions to lower the switching current) has been proposed by Zhu et al.~\cite{Zhu03,Zhu08}.

\section{Experimental Technique}
The nanorings are fabricated by standard electron beam lithography, using a JEOL JSM-7001 F SEM. A double layer MMA/PMMA resist is spin coated on top of a gold-coated silicon wafer substrate.  After e-beam exposure, the samples are developed in a solution of methyl isobutyl ketone and isopropyl alcohol. The desired thickness of cobalt is then deposited by electron beam evaporation, followed by an acetone soak and a sonication ``lift-off" procedure. A thin protective platinum layer $\sim$4~nm is then deposited by thermal evaporation on the ring structure to prevent oxidization of cobalt.

\begin{figure}[tb]
\includegraphics[width=8.5cm]{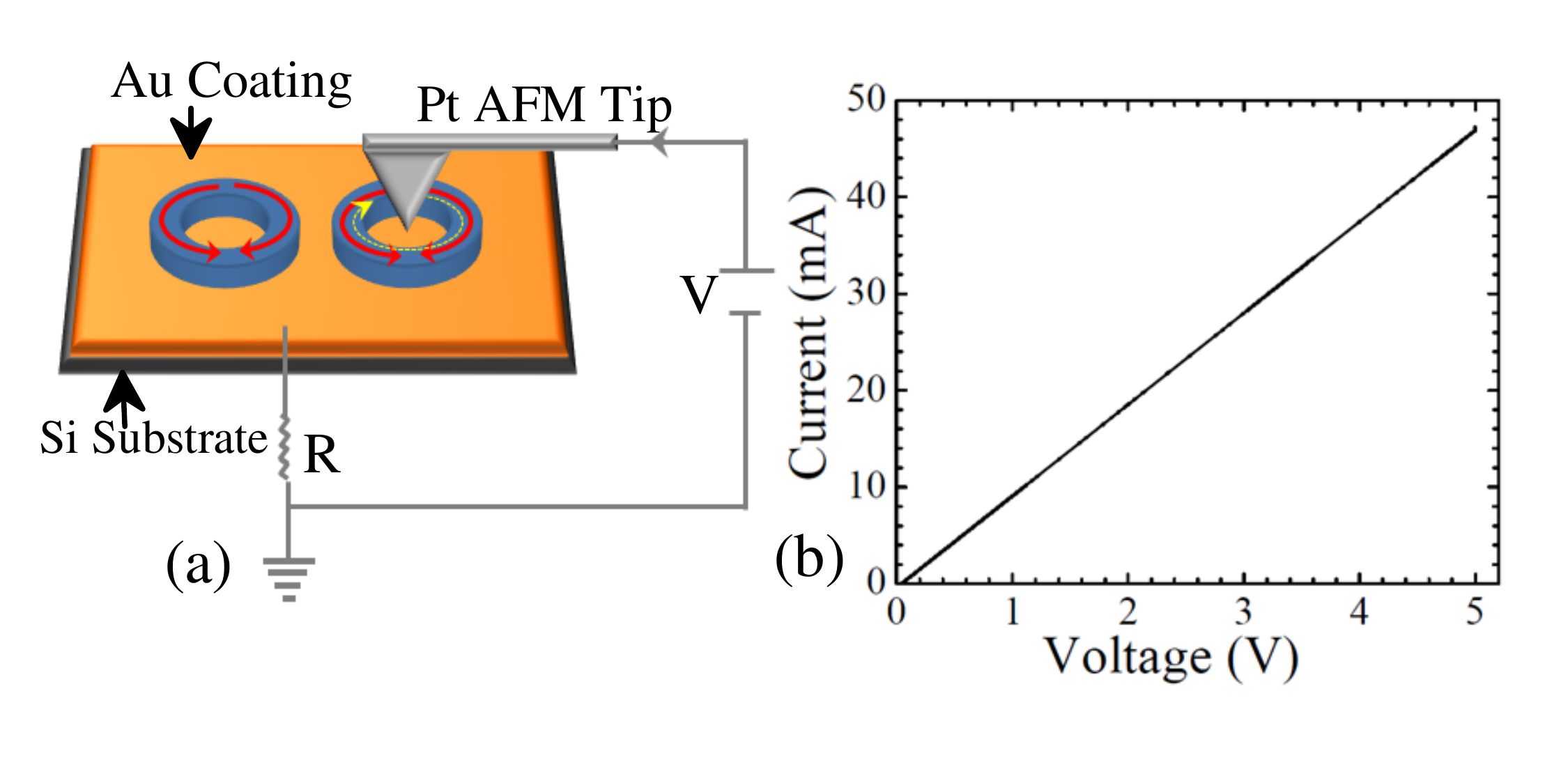}
\caption{\label{Expt}(a) Experimental setup of AFM tip used to pass current through the individual nanoring by contact mode. (b) A current vs voltage graph measured by the experimental setup shown in (a).}
\end{figure}

All AFM images and manipulation are performed with an Asylum Research MFP-3D atomic force microscope. The schematic of the experimental technique is shown in Fig.~\ref{Expt}(a) and can also be found elsewhere~\cite{Pradhan11}. Our procedure consists of three steps.  First, we apply a uniform in-plane field of about $\pm$2500~Oe to the nanoring sample, removing it to obtain the remanent state. We image this state with a low moment magnetic tip (Asylum Research ASYMFMLM). Next, we switch the magnetic tip to a solid platinum metal tip (Rocky Mountain Nanotechnology LLC, tip radius $\sim$25~nm) and locate the same rings imaged with the magnetic tip in the first step.  Finally, we bring the tip into contact with the surface at the center of a ring and pass the desired amount of current through the tip, without contacting the ring arm.  

We measure the voltage across a 100~$\Omega$ resistor in series with the tip to determine the current. A typical current vs. voltage plot is shown in Fig.~\ref{Expt}(b), demonstrating our ability to pass up to 50~mA through the solid metal tip, which we can consistently do 10-20 times without losing our ability to topographically image with the metal tip, despite the high current density.  Passing current into the substrate was taken to be positive, while passing current out of the substrate was take to be negative, with the polarity of the current determining the CW and CCW circulation of the field by the right hand rule.  The strength of the applied circular field can be calculated by Ampere's law using the radius of the ring and the current passed through the tip. Last, we switch back to the magnetic tip to observe the change of the magnetization in the individual ring due to the applied circular field. All measurements were done at room temperature. If the applied circular field was too weak to change the magnetization, then we repeat the second two steps, increasing the applied current until the resulting field is strong enough to change the magnetic state of the nanoring.

\section{Results and Discussion}
\begin{figure*}[bt]
\includegraphics[width=17.0cm]{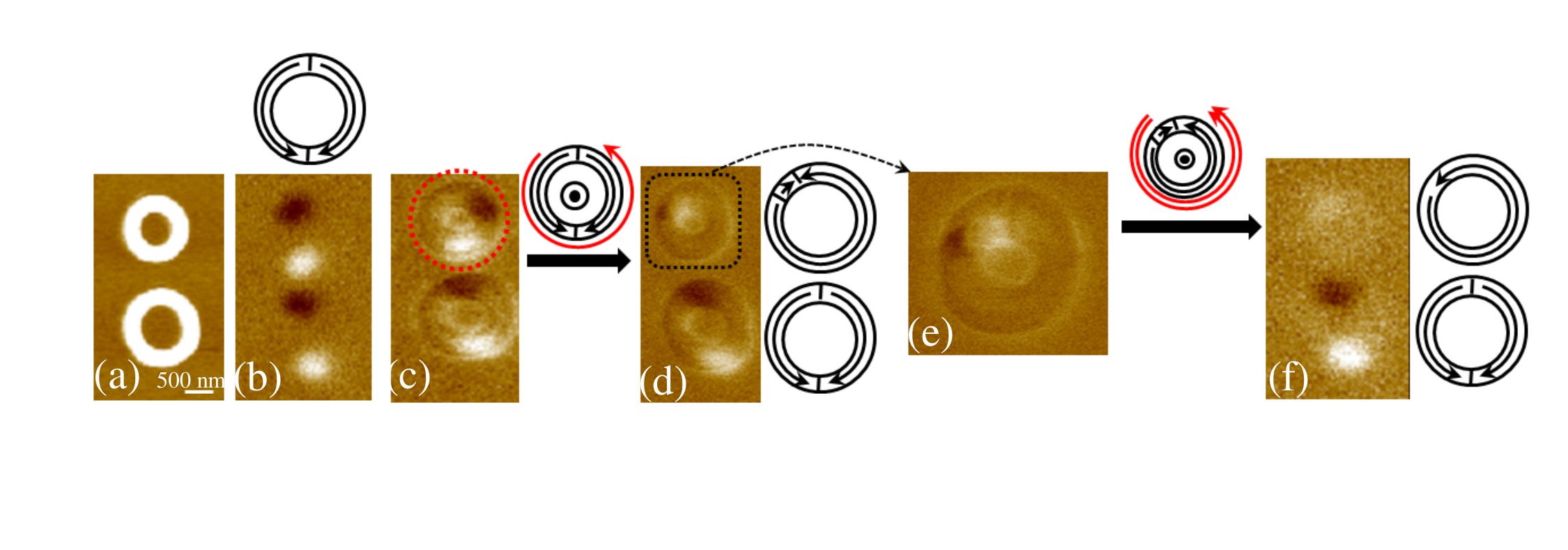}
\caption{\label{Onion_Vortex}(a) AFM height image of two symmetric rings with 900~nm (top) and 1050~nm outer diameter with ring arm 140~nm and thickness 20~nm. (b) MFM image of the rings at 300~Oe in-plane magnetic field showing onion state. Dark and bright contrasts corresponds to tail-to-tail (T-T) and head-to-head (H-H) $180^\circ$ DWs. (c) MFM images when field was reduced to zero, shows the rings are in onion state. Circled ring choosen to apply circular field. (d) MFM image after applied circular field and (e) is the zoomed MFM image of top ring after applied circular field. (f) MFM images of rings after applied stronger circular field on top ring.}
\end{figure*}
Figure~\ref{Onion_Vortex} shows Onion-$360^\circ$DW-Vortex switching under the applied circular field. Figure~\ref{Onion_Vortex}(a) is the height image of two symmetric cobalt nanorings with the same arm width of 140~nm and different outer diameters of 900~nm (top) and 1050~nm (bottom). The nanorings are 20~nm thick. We initialize the rings in a uniform 2500~Oe in-plane field pointing towards the bottom of the image.  The MFM image in Figure~\ref{Onion_Vortex}(b) reveals the onion state in both the rings with 300~Oe in-plane field, with dark contrast and bright contrast corresponding to the T-T and H-H $180^\circ$ DWs. The cartoon on the top of MFM image shows the magnetic configuration of the onion state.

When the applied field is reduced to zero, the DWs shift their positions, shown by the MFM image in Fig.~\ref{Onion_Vortex}(c). We applied the circular field to only the top ring, indicated by the dashed red circle, by passing current through a solid platinum tip as described above.  A CCW field resulted from the -43~mA passed through the center of the ring, corresponding to an applied field of 180~Oe at the ring average circumference  estimated from Ampere's law for an infinite wire. This current corresponds to a current density on the order of $10^9$~A/cm$^2$ assuming a 25~nm tip radius.  The red arrow around the cartoon indicates the direction of applied circular field. Figure~\ref{Onion_Vortex}(d) shows the MFM image of resulting magnetic states of both the rings, in which the larger ring (bottom) remains the same, while in the small ring (top), the two $180^\circ$DWs moved together and combined to form a $360^\circ$DW with vortex chirality along the direction of applied field. The cartoon on the right side of the MFM image [Fig.~\ref{Onion_Vortex}(d)] sketches the magnetic states of the rings. The strength of the field applied to the top ring does not affect the nearby elements, indicated by the lack of change in the bottom ring. Contrast changes are expected due to wear on the metal coated tip that is unavoidable due to the substantial topographic imaging required to locate the identical rings.

A higher resolution MFM image of the top ring (Fig.~\ref{Onion_Vortex}(e)) clearly shows the dark and bright contrast of the $360^\circ$DW (or combined two $180^\circ$DWs). We next increase the field strength by passing a higher current (-48~mA = 225~Oe CCW) through the same ring. The double red arrow around the cartoon indicates the higher circular field strength and direction. Figure~\ref{Onion_Vortex}(f) shows the resulting magnetic states after passing higher current/applying higher field. The top ring now shows no contrast, indicating the annihilation of $360^\circ$DW to form the CCW vortex state. The cartoon on the right of the MFM image shows the magnetic configuration of the rings. Overall, the two $180^\circ$DWs formed initially in the onion state were forced toward one another by the applied circular field to form a $360^\circ$DW and then annihilated by increasing the field strength. This demonstrates the ability to create the $360^\circ$DW, which has been reported by other groups~\cite{Ross03a,Ross03b,Ross04,Zhu03,Muratov09,Pradhan11}, and to annihilate the DWs with a sufficiently strong field.  We next demonstrate the proposed switching mechanism from one vortex circulation to the other, keeping the $360^\circ$DW.

\begin{figure}[ht]
\includegraphics[width=8.5cm]{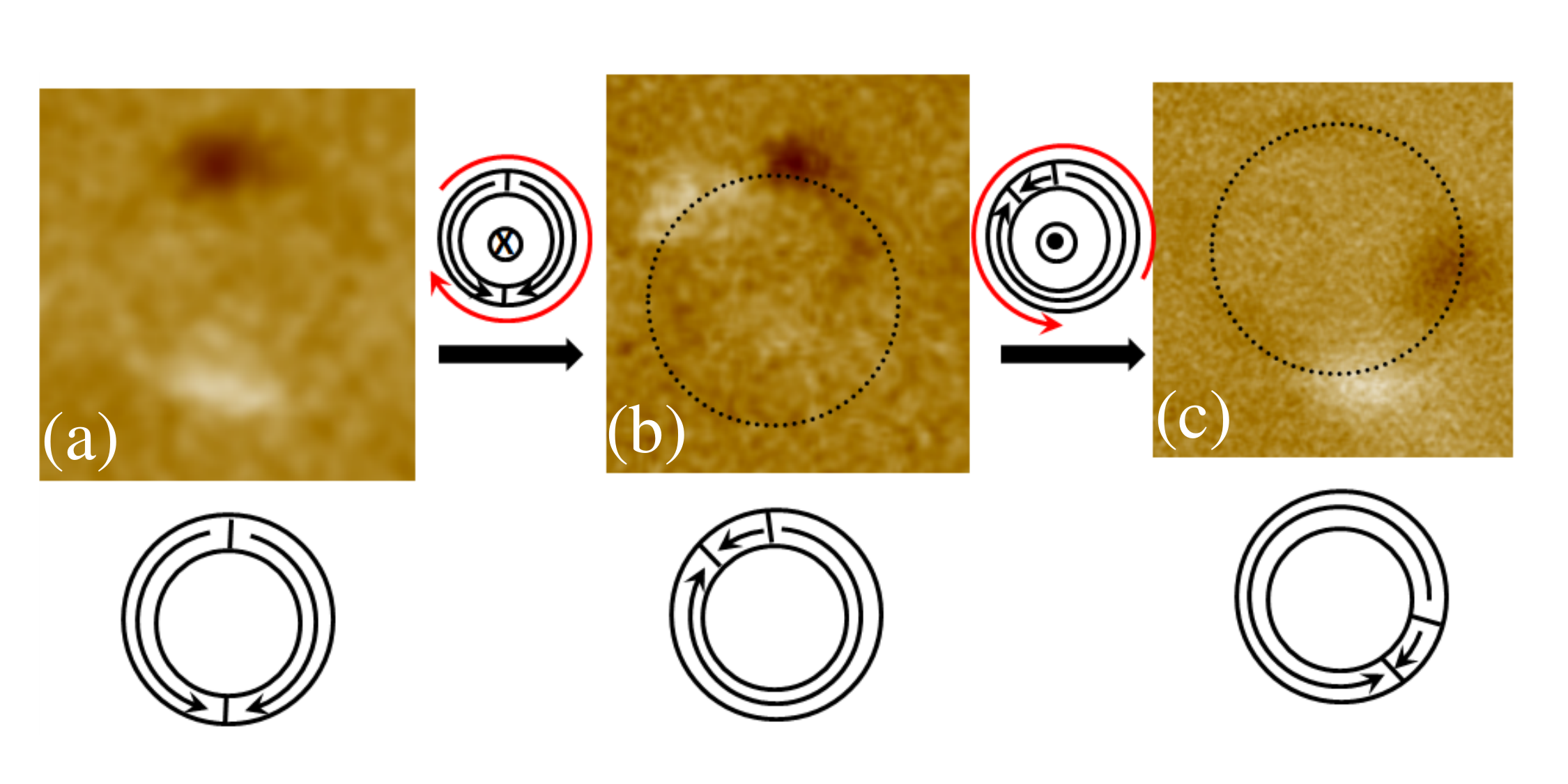}
\caption{\label{Exp-Comb}(a) MFM image of symmetric nanorings with 900~nm outer diameter, 15~nm thickness and width 100~nm, showing the onion state at zero field. (b) MFM image of the rings after a CW circular field was applied. The two $180^\circ$DW move towards one another. (c) MFM images after CCW circular field was applied to the ring. The two $180^\circ$DW split and recombined on the opposite side of the ring arm.}
\end{figure}

\begin{figure}[ht]
\includegraphics[width=8.5cm]{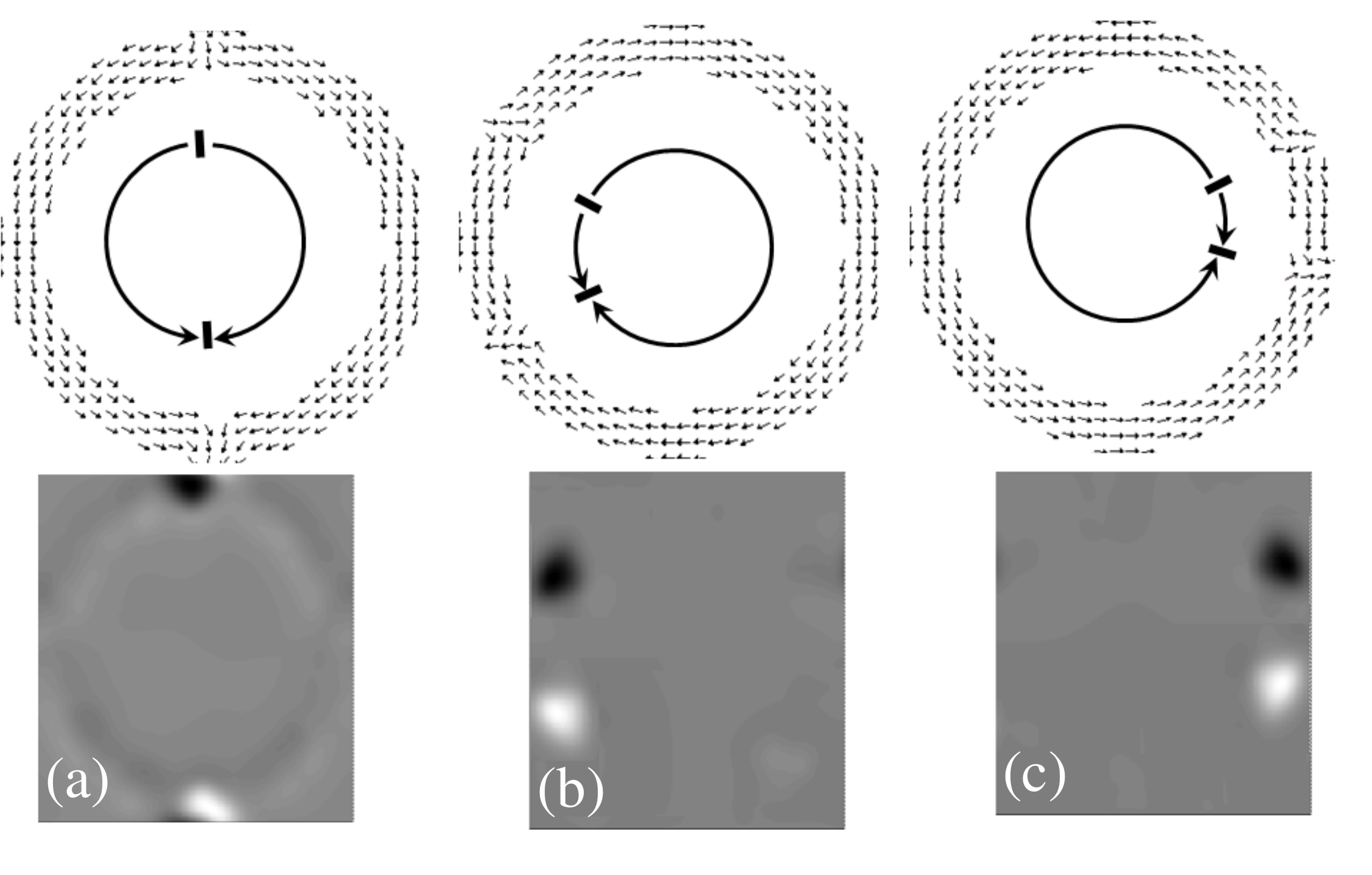}
\caption{\label{sim_combo} Simulated micromagnetic states of the ring in Fig. \ref{Exp-Comb} and corresponding MFM image:  (a) initial onion state (b) CW $360^\circ$ state (c) CCW $360^\circ$ state.}
\end{figure}

Figure~\ref{Exp-Comb} shows the experimental implementation of Onion-CW $360^\circ$DW -CCW $360^\circ$DW switching with an applied circular field. The ring has an outer diameter  900~nm, thickness of 15~nm and width of 100~nm. Figure~\ref{Exp-Comb}(a) shows the initial onion state of the nanoring at remanence, in which the dark and bright contrast indicates the T-T and H-H domain walls, respectively. A CW circular field was applied by passing current (40~mA = 188~Oe) through the center of the ring. Figure~\ref{Exp-Comb}(b) reveals the resulting magnetic state.  The two $180^\circ$DWs moved towards one another forming the twisted or $360^\circ$ state with CW vortex circulation, as controlled by the applied circular field. The T-T DW appears to be pinned and the H-H DW moves near the T-T DW. The pinning could be due to topographical defects in the nanoring through the fabrication and strength of the applied field was insufficient to move the pinned DW (T-T).  The cartoon below the MFM image sketches the magnetic state of the nanoring. Instead of applying a higher field in the same direction as before in Fig.~\ref{Onion_Vortex}(e), we applied the reverse circular field (CCW direction) by changing the direction of the current (-45~mA = -211~Oe) through the center of the ring. Figure~\ref{Exp-Comb}(c) shows the MFM image after the reverse circular field was applied. The two combined $180^\circ$DWs pulled apart, moved to the opposite side of the ring and recombined to form a CCW vortex circulation, again following the direction of the applied field. 

Figure~\ref{sim_combo} is a micromagnetic simulation of a ring with identical geometry as the ring in the experiment, using physical parameters for cobalt carried out with OOMMF, the micromagnetic software package developed by NIST~\cite{OOMMF}. The magnetization states are shown on top, with cartoon inside each figure schematically showing the magnetic states to aid in interpretation.  The corresponding MFM simulation is below each magnetic state.  The onion state is initialized in the same direction as the experiment (Fig.~\ref{sim_combo}a). The two $180^\circ$DWs move symmetrically to the left (Fig.~\ref{sim_combo}b)and right (Fig.~\ref{sim_combo}b) side of the ring arms after experiencing a simulated current of 40~mA, due to a lack of any topographical defects.  The relatively good quantitative agreement in field strength between the experiment and simulations should be taken as fortuitous; the simulations are carried out at 0~K and there is no pinning of the DWs as the ring is taken to be perfectly symmetric and defect free.

\section{Conclusion}
In conclusion, we experimentally demonstrate control over $\pm$$360^\circ$DW magnetic states in thin cobalt nanorings by applying a circular magnetic field generated by passing current through the ring center with an AFM tip. The $360^\circ$DW vortex states can be easily switched back and forth by reversing the direction of the current passed through the ring.  These states are suitable for MRAM devices, as previously proposed~\cite{Zhu03, Muratov09}. We estimate the applied current density to be around $10^9$~A/cm$^2$, which would be smaller if our AFM tip filled the entire center of the ring.  The switching field can be further reduced by decreasing the thickness and width of the ring~\cite{Muratov09}.  In our case the DW motion is due to the circular field and not due to spin torque transfer.  Passing current through the ring material in addition to the center of the ring would further decrease the the critical switching field/depinning field~\cite{Laufenberg06}.

\begin{acknowledgments}
The authors acknowledged the support by NSF grant No. DMR-0906832, the Research Corporation for Science Advancement Grant No.7889, and the NSF Center for Hierarchical Manufacturing (CMMI-0531171).
\end{acknowledgments}


\begin{thebibliography}{99}

\bibitem{Zhu08} J. G. Zhu, Proceeding of the IEEE {\bf 96} 11, 1786, (2008).
\bibitem{Zhu00} J. G. Zhu, Y. F. Zheng and G. A. Prinz, J. Appl. Phys. {\bf 87} 6668, (2000).
\bibitem{Zhu03}  Xiaochun Zhu and Jian-Gang Zhu, IEEE TRANSACTIONS ON MAGNETICS, {\bf39} 5, (2003).
\bibitem{Muratov09} Cyrill B Muratov and V. Osipov IEEE TRANSACTIONS ON MAGNETICS, {\bf45} 8, 3027, (2009).
\bibitem{Laufenberg06} M. Laufenberg, W. Buhrer, D Bedau, P. E. Melchy, M. Klaui, L. Vila, G. Faini, C. A. F. Vaz, J. A. C. Bland and U. Rudiger, Phys. Rev. Lett. {\bf 97} 046602, (2006).
\bibitem{Klaui03} M. Klaui, C. A. F. Vaz, L. Lopez-Diaz and J. A. C. Bland, J. Phys.: Condens. Matter, {\bf 15} R985-R1023, (2003).
\bibitem{Klaui07} C. A. F. Vaz1, T. J. Hayward, J. Llandro, F. Schackert, D. Morecroft, J. A. C. Bland, M. Klaui, M. Laufenberg, D. Backes, U. Rudiger,
F. J. Castano, C. A. Ross, L. J. Heyderman, F. Nolting, A. Locatelli,
G. Faini, S. Cherifi and W. Wernsdorfer, J. Phys.: Condens. Matter, {\bf 19} 255207, (2007).
\bibitem{Zhu06} F. Q. Zhu, G. W. Chern, O. Tchernyshyov, X. C. Zhu, J. G. Zhu and C. L. Chein, Phys. Rev. Lett. {\bf 96} 027205, (2006).
\bibitem{Saitoh04} E. Saitoh, M. Kawaba, K. Harii and H. Miyajima, J. Appl. Phys. {\bf 95} 1986, (2004).
\bibitem{Pradhan11} T. Yang, Nihar R. Pradhan, A. Goldman, A. Licht, Y. Li, M. Kamei, M. T. Tuominen and K. E. Aidala, Appl. Phys. Lett. {\bf 98}, 242505 (2011).
\bibitem{OOMMF}
{OOMMF is available for free from NIST at http://math.nist.gov/oommf}.
\bibitem{Gabriel09} Gabriel D. Chaves-O'Flynn, A. D. Kent and D. L. Stein, Phys. Rev. B. {\bf 79} 184421 (2009).
\bibitem{Ross03a} F. J. Castano, C. A Ross and A. Eilez, J. Phys. D: Appl. Phys {\bf36} 2031, (2003).
\bibitem{Ross03b} F. J. Castano, C. A. Ross, C. Frandsen, A. Eilez, I. Smith, M. Redjdal and F. B. Humphrey, Phys. Rev. B. {\bf67} 184425, (2003).
\bibitem{Ross04} F. J. Castano, C. A. Ross, A. Eilez, W. Jung and  C. Frandsen, Phys. Rev. B. {\bf69} 144421, (2004).

\end{thebibliography}
\end{document}